\documentclass[prb,twocolumn,nofootinbib,amssymb]{revtex4-1}
\usepackage{natbib}
\usepackage{graphicx}
\usepackage{dcolumn}
\usepackage{amsmath,amssymb,amsthm}
\usepackage{color}
\begin{document}

\title{Quantum turnstile regime of nanoelectromechanical systems}

\author{R. Dragomir}
\affiliation{National Institute of Materials Physics, Atomistilor 405A, Magurele 077125, Romania}
\author{V. Moldoveanu}
\affiliation{National Institute of Materials Physics, Atomistilor 405A, Magurele 077125, Romania}
\author{S. Stanciu}
\affiliation{Faculty of Physics, University of Bucharest, Atomistilor 405, Magurele 077125, Romania}
\affiliation{National Institute of Materials Physics, Atomistilor 405A, Magurele 077125, Romania}
\author{B. Tanatar}
\affiliation{Department of Physics, Bilkent University, Bilkent, 06800 Ankara, Turkey}

\newcommand{\VM}{\textcolor{red}}
\newcommand{\BT}{\textcolor{blue}}
\newcommand{\RD}{\textcolor{magenta}}
\newcommand{\BS}{\textcolor{green}}

\begin{abstract}

The effects of a turnstile operation on the current-induced vibron dynamics in nanoelectromechanical 
systems (NEMS) are analyzed in the framework of the generalized master equation. In our simulations each turnstile
cycle allows the pumping of up to two interacting electrons across a biased mesoscopic subsystem which is 
electrostatically coupled to the vibrational mode of a nanoresonator. The time-dependent mean vibron number 
is very sensitive to the turnstile driving, rapidly increasing/decreasing along the charging/discharging 
sequences. This sequence of heating and cooling cycles experienced by the nanoresonator is due to  specific 
vibron-assisted sequential tunneling processes along a turnstile period. At the end of each charging/discharging 
cycle the nanoresonator is described by a linear combination of vibron-dressed states $s_{\nu}$ associated 
to an electronic configuration $\nu$. If the turnstile operation leads to complete electronic depletion the 
nanoresonator returns to its equilibrium position, i.e.\,its displacement vanishes. It turns out that a suitable 
bias applied on the NEMS leads to a slow but complete cooling at the end of the turnstile cycle. Our calculations 
show that the quantum turnstile regime switches the dynamics of the NEMS between vibron-dressed subspaces
with different electronic occupation numbers. We predict that the turnstile control of the electron-vibron 
interaction induces measurable changes on the input and output transient currents.     

\end{abstract}

\maketitle

\section{Introduction}

The nanoelectromechanical systems are hybrid structures in which the electrostatic interaction between 
vibrational modes and open mesoscopic systems is expected to play a role down to the quantum level \cite{Poot}.
To support this idea, the sensing properties of nanoresonators (NR) in the presence of electronic transport have been 
investigated in various experimental settings. 

For instance, singly clamped cantilevers or AFM tips were shown to record single-electron tunneling from back-gate contacts to 
the excited states of quantum dots deposited on a substrate \cite{Cockins}. In another class of experiments, a suspended 
carbon nanotube (CNT) with an embedded quantum dot is actuated by microwave signals and the dips of its resonance frequency are 
associated to single-electron tunneling \cite{Meerwaldt}. Besides flexural modes, the CNTs also develop longitudinal modes with 
higher frequencies (up to few GHz). Similarly, the vibration energy $\hbar\omega$ of single-molecule junctions is around few meVs \cite{Franke}. 
For these systems, refined cooling techniques were used to reach the regime $\hbar\omega\gg k_BT$ for 
which the vibrations of the nanoresonator must be quantized \cite{LaHaye,O'Connel}.   

On the other hand, the implementation of nanoelectromechanical systems as successful devices in quantum sensing \cite{Degen}, 
molecular spintronics or nano-optomechanics \cite{Aspelmeyer} requires an accurate tuning of the underlying electron-vibron coupling. 
For example, the electron-vibron coupling can be switched on and off by controlling the {\it location} of a quantum dot (QD) along 
the suspended CNT in which it is formed \cite{Benyamini,Weber}. 

\begin{figure}[t]
 \includegraphics[angle=0,width=0.3\textwidth]{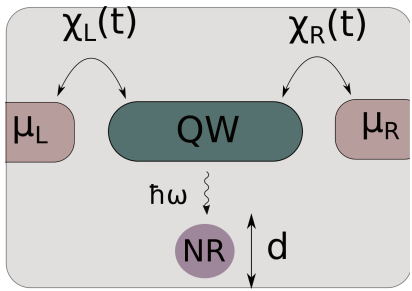}
	\caption{Schematic view of the NEMS in the turnstile regime. Source ($L$) and drain ($R$) 
	particle reservoirs with chemical potentials $\mu_{L,R}$ are connected to an electronic structure 
	(e.g.\, a quantum wire - QW). The contact regions are modulated by switching functions $\chi_{L,R}$ - the turnstile 
	operation corresponds to periodic out-of-phase oscillations of $\chi_{L,R}$. 
	A vibrational mode of frequency $\omega$ interacts with the electrons, $d$ being its displacement 
	w.r.t\, the equilibrium position.}
 \label{fig1-NEMS}
 \end{figure}

In this theoretical study we focus on the time-dependent control of the entangled electron-vibron dynamics of a 
NEMS in the quantum turnstile regime. More precisely, we show that the pumping of an 
integer number of electrons along a turnstile period activates the coupling to the vibrational mode during the charging cycle 
and then renders it ineffective on the discharging cycle when the system is fully depleted. We recall that in the turnstile setup 
\cite{TSP1,TSP2,TSP3}, electrons are first injected from the source (left) particle reservoir while the contact to 
the drain (right) reservoir is closed. After this charging half-period, the left/right contact closes/opens simultaneously 
(see the sketch in Fig.\,\ref{fig1-NEMS}).

In most experimental investigations on NEMS, a bias voltage continuously supplies the charge flow through 
the mesoscopic system which in turn interacts with the vibrational mode. Then the hybrid structure evolves under the 
electron-vibron coupling until a stationary transport regime is reached. At the theoretical level, the latter is 
recovered by solving rate equations \cite{Galperin,Erpenbeck2} or hierarchical quantum master equations (HQME) \cite{Schinabeck}. 
Also, the single-level Anderson-Holstein model provides a sound description of the essential spectral properties of NEMS via 
the Lang-Firsov polaron transformation. 

Let us stress that recent observation of real-time vibrations in CNTs \cite{Barnard,Gotz} and pump-and-probe measurements \cite{Khivrich} 
provide a strong motivation to scrutinize the time-dependent vibron-assisted transport. Few theoretical descriptions of vibron-assisted 
transport properties in the presence of pumping potentials acting on the electronic system can be mentioned. The effect of a cosine-shaped 
driving of the contact regions has been considered within the Floquet Green's function formalism \cite{Haughian1,Haughian2}.
In a very recent paper the HQME method was adapted for a time-dependent setting \cite{Erpenbeck}.
Avriller {\it et al.} \cite{Avriller} calculated the transient vibron dynamics induced by a step-like coupling of molecular 
junctions to source and drain particle reservoirs.

In the present work we rely on the generalized master equation (GME) method which was previously used to study the turnstile 
regime of single-molecule magnets \cite{NJP-SMM} and recently extended for hybrid systems such as NEMS or cavity-QD systems \cite{Entropy}. 
The model Hamiltonian embodies both the electron-electron interaction within the electronic subsystem and the spin degree of freedom.
We also consider turnstile operations where more than one electron is transferred across the system. 
The reduced density operator of the hybrid system is calculated numerically with respect to vibron-dressed basis. 
As we are interested in the response of the NR to the turnstile pumping we also calculate its 
associated displacement which can be, in principle, measured. Note that this quantity is mostly derived for the classical regime of 
nanoresonators via the Langevin equation \cite{Hussein}. 

The rest of the paper is organized as follows. In Section II we introduce the model and 
briefly recall the main ingredients of the GME approach. The results are presented in Section III, Section IV being left 
to conclusions.     


\section{Formalism}

A typical NEMS setup is sketched in Fig.\,\ref{fig1-NEMS} where a quantum wire (QW) is capacitively coupled to a nearby
nanoresonator (NR) and tunnel-coupled to source and drain leads. The closed nanoelectromechanical system (i.e.\,not 
connected to particle reservoirs) is described by the following general Hamiltonian
\begin{equation}\label{H-system}
H_S=H_{S,0}+V_{{\rm el-vb}},
\end{equation}
where $H_{S,0}$ accounts for the two components of the NEMS, i.e.\,the QW accommodating several interacting electrons 
and the vibrational mode with frequency $\omega$ associated to a molecule or 
a nanoresonator:
\begin{eqnarray}\nonumber
	H_{S,0}&=&\sum_{i,\sigma}\varepsilon_{i\sigma}c^{\dagger}_{i\sigma}c_{i\sigma}+
       \frac{1}{2}\sum_{\sigma,\sigma'}\sum_{i,j,k,l}V_{ijkl}c^{\dagger}_{i\sigma}c^{\dagger}_{j\sigma'}c_{l\sigma'}c_{k\sigma}\\\label{H_S0}
	&+&\hbar\omega a^{\dagger}a.
\end{eqnarray}
Here $c^{\dagger}_{i\sigma}$ creates an electron with spin $\sigma$ on the single-particle state $\psi_{i\sigma}$ 
of the electronic system with the corresponding energy $\varepsilon_{i\sigma}$, the second term is the 
two-body Coulomb interaction 
within the electronic sample and $a^{\dagger}$ is the creation operator for vibrons. The eigenstates $|\nu,N\rangle$ 
of $H_{S,0}$ are products of electronic many-body configurations $|\nu\rangle$ with energies $E_{\nu}$ of the 
electronic system and $N$-vibron Fock states $|N\rangle$, such that 
$H_{S,0}|\nu,N\rangle=(E_{\nu}+N\hbar\omega )|\nu,N\rangle$. The electron-vibron coupling $V_{{\rm el-vb}}$ reads as
\begin{equation}\label{V-el-bin}
V_{{\rm el-vb}}=\sum_{i,\sigma}\lambda_ic^{\dagger}_{i\sigma}c_{i\sigma}(a^{\dagger}+a),
\end{equation}	
where $\lambda_i$ is the electron-vibron coupling strength. 

We denote by ${\cal E}_{\nu,s}$ and $|\varphi_{\nu,s}\rangle$ the eigenvalues and eigenfunctions of the hybrid system such that
\begin{equation}\label{Eigenvalues}
H_S|\varphi_{\nu,s}\rangle={\cal E}_{\nu,s}|\varphi_{\nu,s}\rangle.
\end{equation}
Since $V_{{\rm el-vb}}$ conserves the electronic occupation and the spin, the fully interacting states 
$|\varphi_{\nu,s}\rangle$ can still be labeled by a many-body configuration $\nu$ and written as:
\begin{equation}\label{full-states}
|\varphi_{\nu,s}\rangle=|\nu\rangle\otimes\left\lbrace\sum_N A^{(\nu)}_{sN}|N\rangle\right\rbrace:
=|\nu\rangle\otimes |s_{\nu}\rangle.
\end{equation}
The $\nu$-dependent vibrational overlap $|s_{\nu}\rangle$ contains different states $|N\rangle$, $A^{(\nu)}_{sN}$ being the weight of 
the $N$-vibron state. If $|\varphi_{\nu,s}\rangle$ are obtained by 
numerical diagonalization one should truncate the indices $N$ and $s$ at a convenient upper bound $N_0$. 
In this case, the coefficients $A^{(\nu)}_{sN}$ define a finite dimensional unitary matrix which approximates the exact Lang-Firsov 
transformation defined by the operator $S=\sum_{i,\sigma}(\lambda_i/\hbar\omega) c^{\dagger}_{i\sigma}c_{i\sigma}(a^{\dagger}-a)$ 
(see e.g., Ref.\,\onlinecite{Thoss1}).
The exact eigenfunctions are then $|\varphi_{\nu,s}\rangle=e^{-S}|\nu,N\rangle$. 

Let us stress that the electron-vibron coupling constants $\lambda_i$ depend on the single-particle wavefunctions 
$\psi_{i\sigma}$ of the electronic subsystem. In a recent work \cite{NEMS-1} we took this dependence into account and showed 
that it leads to different sensing efficiencies when a singly-clamped tip is placed above the quantum wire and swept along it. 
In this work the position of the NR is fixed and the transport involves, for simplicity, only the lowest spin-degenerate 
single-particle state whose associated electron-vibron coupling strength will be denoted by $\lambda_0$. 
It is useful to introduce the Franck-Condon factors (FC):
\begin{equation}\label{FC-factors}
	F_{\nu\nu';ss'}:=\langle s_{\nu}|s'_{\nu'}\rangle=\sum_{N=0}\overline {A^{(\nu)}_{sN}}A^{(\nu')}_{s'N},\quad |n_{\nu}-n_{\nu'}|=1,
\end{equation}
where $n_{\nu}$ is the number of electrons corresponding to the many-body configuration $\nu$. We shall see below that for a 
given pair of electronic configurations $\{\nu,\nu'\}$ one gets a series of vibron-assisted transitions controlled by $F_{\nu\nu';ss'}$.

In view of vibron-assisted transport the electronic component of NEMS is also coupled to source (L) and drain (R) 
particle reservoirs characterized by chemical potentials $\mu_{L,R}$, as shown in Fig.\,\ref{fig1-NEMS}. 
The total Hamiltonian therefore becomes:
\begin{equation}\label{Htotal}
	H(t)=H_S+\sum_{l=L,R}H_l+H_T(t),
\end{equation}
where $H_l$ is the Hamiltonian of the lead $l$ and the tunneling Hamiltonian reads as ($h.c.$ denotes Hermitian conjugate):
\begin{equation}\label{Htunnel}
	H_T(t)=\sum_{l=L,R}\sum_{i,\sigma}\int dq \chi_l(t)\left (T^{(l\sigma)}_{qi}c^{\dagger}_{ql\sigma}c_{i\sigma}+h.c. \right).
\end{equation}
The functions $\chi_l(t)$ simulate the turnstile modulation of the contact barriers between the leads and the system and
$T^{(l\sigma)}_{qi}$ is the coupling strength associated to a pair of single-particle states from the lead $l$ and the central 
sample. For simplicity we assume that the tunneling processes are spin conserving and that $T^{(l\sigma)}_{qi}$ does not depend 
on $\sigma$. We describe the leads as one-dimensional semi-infinite discrete chains which feed both spin up and down electrons 
to the central system. Their spectrum is $\varepsilon_{q_l}=2t_L\cos q_l$, where $q_l$ is electronic momentum in the lead $l$ 
and $t_L$ denotes the hopping energy on the leads.

The reduced density operator (RDO) $\rho$ of the hybrid system obeys a generalized master equation (GME)
(for a derivation via the Nakajima-Zwanzig projection method see e.g., Ref.\,\onlinecite{Entropy}):
\begin{eqnarray}\nonumber
	&&\frac{\partial\rho(t)}{\partial t}=-\frac{i}{\hbar}[H_S,\rho(t)]-(n_B+1){\cal L}_{\kappa}[a]\rho(t)\\\label{GME}
	&-&n_B{\cal L}_{\kappa}[a^{\dagger}]\rho(t)-\frac{1}{\hbar^2}\int_{t_0}^tds {\rm Tr}_L 
	\left\lbrace {\cal K}(t,t-s;\rho(s)) \right\rbrace,
\end{eqnarray}
where ${\rm Tr}_L$ is the partial trace with respect to the leads' degrees of freedom and we introduced the non-Markovian dissipative kernel 
due to the reservoirs:
\begin{equation}\label{K-leads}
	{\cal K}(t,t-s;\rho(s)):=\left [H_T(t),U_{t-s}[H_T(s),\rho(s)\rho_L ]U_{t-s}^{\dagger}\right ]. 
\end{equation}
The right hand side of Eq.\,(\ref{GME}) also contains Lindblad-type operators which capture the effect of a thermal bath 
described by the Bose-Einstein distribution $n_B$ and by the temperature $T$ ($\kappa$ 
is the loss parameter) \cite{Utami}: 
\begin{equation}\label{kappa1}
	{\cal L}_{\kappa}[a]\rho(t)=\frac{\kappa}{2}\left (a^{\dagger}a\rho+\rho a^{\dagger}a-2a\rho a^{\dagger}\right ).
\end{equation}
In Eq.\, (\ref{K-leads}) $U_{t}=e^{-\frac{i}{\hbar}(H_S+H_L+H_R)t}$ is the unitary evolution of the disconnected systems 
(i.e\, NEMS+leads). Also, $\rho_L$ is the equilibrium density operator of the leads.

The GME is solved numerically with respect to the vibron-dressed basis  $\{\varphi_{\nu,s}\}$ of the hybrid system. Let us stress 
that choosing the fully interacting basis over the `free' one $\{|\nu,N\rangle\}$ allows us to calculate the 
matrix elements of $e^{\frac{i}{\hbar}H_St}c^{\dagger}_{i\sigma}e^{-\frac{i}{\hbar}H_St}$ which appear in the dissipative 
kernel of the leads  (see Eq.\, (\ref{K-leads})). Note also that in this representation the Lang-Firsov transformation of 
the tunneling Hamiltonian is not needed such that $H_T$ does not acquire an additional operator-valued exponential. 
By doing so one carefully takes into account the FC factors which can have both positive and negative signs, as 
pointed out in Ref.\,\cite{Hubener}.

The full information on the system dynamics is embodied in the populations of various states
\begin{equation}
P_{\nu,s}(t)=\langle\varphi_{\nu,s}|\rho(t)|\varphi_{\nu,s}\rangle.
\end{equation}
The time-dependent currents in each lead are identified from the continuity equation of the charge occupation ${\cal Q}_S$ 
of the system:
\begin{equation}
	\frac{d}{dt}{\cal Q}_S(t)=e{\rm Tr}_{\varphi}\left\lbrace \hat{N}_S\frac{d}{dt}\rho(t)\right\rbrace=J_L(t)-J_R(t),
\end{equation}
where $\hat{N}_S=\sum_{i,\sigma}c^{\dagger}_{i\sigma}c_{i\sigma}$ is the particle number operator, ${\rm Tr}_{\varphi}$ 
stands for the trace with respect to the basis $\{\varphi_{\nu,s}\}$ of the hybrid system and $e$ is the electron charge. 
The left and right transient currents $J_{L,R}$ are then
calculated by collecting all diagonal elements $\langle\varphi_{\nu,s}|{\dot \rho}(t)\hat{N}_S|\varphi_{\nu,s}\rangle$
which contain the Fermi function $f_{l=L,R}$. The latter appears when performing the partial trace of the integral kernel 
${\cal K}(t,t-s;\rho(s))$ such that ${\rm Tr}_L \left\lbrace \rho_L c^{\dagger}_{q'l'\sigma'}c_{ql\sigma}\right\rbrace
=\delta_{ll'}\delta_{\sigma\sigma'}\delta(q-q')f_l(\varepsilon_{q_l})$  Also, note that from the cyclic property of 
the trace one has ${\rm Tr}_{\varphi}\{[H_S,\rho(t)]\hat{N}_S\}={\rm Tr}_{\varphi}\{\rho(t)[\hat{N}_S,H_S]\}=0$ and 
${\rm Tr}_{\varphi}\{{\cal L}_{\kappa}[a]\rho(t)\hat{N}_S\}=0$.

Other relevant observables are the average vibron number $N_v={\rm Tr}_{\varphi}\{\rho(t)a^{\dagger}a\}$ and the nanoresonator displacement 
\begin{equation}\label{displ}
d=l_0{\rm Tr}_{\varphi}\{(a^{\dagger}+a)\rho(t)\},
\end{equation}
where $l_0=\sqrt{\frac{\hbar}{2M\omega}}$
is the oscillator length and $M$ is the mass of the nanoresonator. 

\section{Numerical results and discussion}

The nanoelectromechanical system considered in our calculations is made of a two-dimensional quantum nanowire 
connected to source and drain reservoirs and a vibrational mode. The latter describes either a nearby suspended CNT 
which supports longitudinal stretching modes or a vibrating molecule deposited on a substrate. The length and width of the nanowire 
are $L_x=75$\,nm and $L_x=15$\,nm, while for the mass of the nanoresonator 
we set $M=2.5\times 10^{-15}$\,kg. The turnstile operation is switched-on at instant $t_0=0$. 
The bias applied on the system is given by $eV=\mu_L-\mu_R$.

\subsection{Vibron-dressed states and tunneling}

In the following we express the lowest two single-particle energies of the 
conducting system with respect to the equilibrium chemical potential of the leads $\mu_0$. Specifically, 
$\varepsilon_{1\sigma}=0.875$\,meV 
and $\varepsilon_{2\sigma}=3.875$\,meV. We choose $t_L=2$\,meV and the vibron energy $\hbar\omega=0.329$\,meV which is in 
the range of the observed longitudinal stretching modes of CNTs \cite{Weber}. The value of the 
electron-vibron coupling parameter is $\lambda_0=0.096$\,meV. The temperature of the particle reservoirs equals that of the 
thermal bath. We chose $k_BT=4.3\,\mu$eV which corresponds to a temperature of 50\,mK. 

The Hamiltonian $H_S$ of the hybrid system is diagonalized within a truncated subspace containing `free' states
$|\nu,N\rangle$ obtained from the lowest-energy 16 electronic configurations and up to $N_0=15$ vibronic states. 
In the presence of electron-vibron coupling one gets an $N_0$-dimensional vibronic manifold $\{\varphi_{\nu,s}\}_{s=0,...N_0}$ 
associated to each electronic configuration $|\nu\rangle$. 

For simplicity we set the chemical potentials of the leads $\mu_{L,R}<\varepsilon_{2\sigma}$ such that the tunneling 
processes involve only the lowest energy one- and two-particle configurations. Then only four electronic configurations 
 will contribute to the transport, namely the empty state $|0\rangle$, two spin-degenerate 
single-particle states $|\uparrow_1\rangle ,\downarrow_1\rangle$ and the two-electron ground state $|\uparrow_1\downarrow_1\rangle$. 
Henceforth we shall drop the level index and use $\uparrow,\downarrow$ instead of $\uparrow_1,\downarrow_1$. 

The transport through the hybrid system is then due to the states $|\varphi_{0,s}\rangle$, $|\varphi_{\uparrow,s}\rangle$, 
$|\varphi_{\downarrow,s}\rangle$ and $|\varphi_{\uparrow\downarrow,s}\rangle$. Clearly, $|\varphi_{0,s}\rangle=|0,s\rangle$ such 
that $s$ is simply the vibron number of a Fock state, because the electron-vibron coupling does not change the `empty' states. 
For a mixed vibrational state $|\varphi_{\nu\neq 0,s}\rangle$, $s$ is related to the integer part of its corresponding 
vibron number $w_{\nu,s}$.  
Indeed, using the Lang-Firsov transformation one obtains the analytical result
\begin{eqnarray}\nonumber
   w_{\nu,s}&=&\langle\varphi_{\nu,s}|a^{\dagger}a|\varphi_{\nu,s}\rangle=
        \langle \nu ,s|e^S a^{\dagger}ae^{-S}|\nu ,s \rangle \\\label{vib-num}
        &=&s+\left (\frac{\lambda_0}{\hbar\omega}n_{\nu} \right )^2,
\end{eqnarray}
where we used the identities  $e^S a e^{-S}=a+\frac{\lambda_0}{\hbar \omega}\hat{N}_S$, 
$\hat{N}_S|\nu ,s \rangle=n_{\nu}|\nu ,s \rangle$ and the fact that $\langle \nu ,s|a^{\dagger}+a|\nu ,s \rangle=0$.
On the other hand, the numerical diagonalization provides $w_{\nu,s}=\sum_{N=0}^{N_0}N|A^{(\nu)}_{sN}|^2$ which fits well to 
Eq.\,(\ref{vib-num}), at least for the lowest vibronic components.

The vibrationally `excited' states correspond to $s>0$, but it should be mentioned that even the lowest-energy states $|\varphi_{\nu,s=0}\rangle$ 
have a non-vanishing vibron number $w_{\nu,0}$ as they are not entirely made of a `free' state $|\nu,N=0\rangle$. Indeed, 
for the parameters considered here we find (see Eq.\,(\ref{full-states})) that the weights of $|\nu,N=0\rangle$ for the one- and 
two-particle states are $|A^{(\uparrow)}_{0,0}|^2=|A^{(\downarrow)}_{0,0}|^2=0.9$ and $|A^{(\uparrow\downarrow)}_{0,0}|^2=0.7$ 
while the corresponding vibron numbers are $w_{\uparrow,0}=w_{\downarrow,0}\approx 0.085$ and $w_{\uparrow\downarrow,0}=4w_{\sigma,0}$. 

Note that the two-particle ground state carries more vibrons 
because the coupling between the conducting system and the NR increases with the particle number.  
Moreover, the diagonal matrix elements of the displacement operator are found as:
\begin{equation}\label{displ-diag}
	d_{\nu}:=\langle\varphi_{\nu,s}|a^{\dagger}+a|\varphi_{\nu,s}\rangle=\frac{2\lambda_0}{\hbar\omega}n_{\nu},
\end{equation}
and therefore depends only on $n_{\nu}$.

In view of transport calculations let us denote by $\Delta_{N,N+1}(s,s')={\cal E}_{\nu,s}-{\cal E}_{\nu',s'}$ the energy 
required to add one electron from the leads such that the hybrid system evolves from an $N$-electron state 
$|\varphi_{\nu',s'}\rangle$ to
the $(N+1)$-electron state $|\varphi_{\nu,s}\rangle$. We calculate these energies for all pairs of configuration 
$\{\nu,\nu'\}$ with a non-vanishing tunneling coefficient 
${\cal T}^{(l\sigma)}_{\nu\nu';ss'}=\langle\varphi_{\nu,s}|c_{\sigma}^{\dagger}|\varphi_{\nu',s'}\rangle f_l({\cal E}_{\nu,s}-{\cal E}_{\nu',s'})$ 
which describes the tunneling-in processes from the $l$-th lead. The tunneling coefficient 
${\cal T}^{(l\sigma)}_{\nu\nu';ss'}$ appears naturally in the Lindblad version of the generalized master equation (see for example
Ref.\,\cite{NEMS-1}) and controls the transport processes
in the quasistationary regime, that is when the charge occupation and mean vibron number do not
depend on time. The argument of the Fermi function reveals the fact that in the quasistationary
regime the energy $\varepsilon_{q_l}$ of the electron entering the sample matches the difference
${\cal E}_{\nu,s}-{\cal E}_{\nu',s'}$ between two configurations of the latter. 
Note that the tunneling amplitudes ${\cal T}^{(l\sigma)}_{\nu\nu';ss'}$
are controlled by the FC factors $F_{\nu\nu\;ss'}$ (see Eq.\,(\ref{FC-factors})).
The same energy differences are relevant for tunneling-out processes
$|\varphi_{\nu,s}\rangle\to |\varphi_{\nu',s'}\rangle$ which are controlled by the  
${\overline f}_l(x)=1-f_l(x)$.

Now, let us discuss the energy differences $\Delta_{N,N+1}(s,s')$ in terms of the difference $\delta=s-s'$. For tunneling-in processes 
one has $\delta>0$ if electrons have enough energy to excite 
more vibrons while for $\delta<0$ the vibrations of the hybrid system are absorbed and allow tunneling of electrons 
from the leads at lower energies. The role of these transitions changes in the case of tunneling-out processes: 
the system is `heated' for $\delta<0$ and `cooled' down if $\delta>0$. On the other hand, from Eq.\,(\ref{vib-num}) 
one notices that if $\lambda_0/\hbar\omega\ll 1$ the average vibron number are only slightly changed by the `diagonal' 
processes $s=s'$.   

\begin{figure}[tbhp!]
\includegraphics[width=0.45\textwidth]{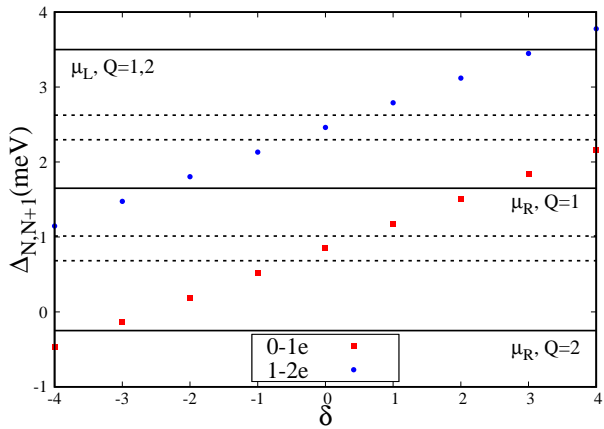}
\caption{The energy differences associated to sequential tunneling processes leading to transitions between electronic
        configurations with $N$ and $N+1$ electrons. $\delta$ denotes the difference between the average number
        of vibrons for the vibrational states $|s_{\nu}\rangle,|s'_{\nu'}\rangle$. For the simplicity of writing we do not
indicate the pairs $(s,s')$ corresponding to the same tunneling energy. The horizontal lines mark the values of the
        chemical potential for which one obtains various turnstile regimes - see the discussion in the text.}
\label{fig2-NEMS}
\end{figure}

Figure\,\ref{fig2-NEMS} displays the tunneling energies as a function of $\delta$ and helps us to identify which transitions 
contribute to the current for a symmetric bias window set by $\mu_{L,R}=E_{\nu}-E_{\nu'}\pm p\hbar\omega/2$, where $p$ is an 
odd positive integer. For example, the four dashed lines in Fig.\,(\ref{fig2-NEMS}) correspond to 
$\mu_{L,R}=\Delta_{0,1}\pm \hbar\omega/2$ and $\mu_{L,R}=\Delta_{1,2}\pm \hbar\omega/2$. 

In agreement with the analytical results obtained via the Lang-Firsov transformation, the eigenvalues corresponding to 
the same electronic configuration $\nu$ are separated by integer multiples of vibron quanta, that is 
${\cal E}_{\nu,s}={\cal E}_{\nu,0}+s\hbar\omega$. This implies that the tunneling energies are also equally spaced, 
that is $\Delta_{0,1}(s,s')={\tilde \varepsilon}_1+(s-s')\hbar\omega$ and
$\Delta_{1,2}(s,s')={\tilde\varepsilon}_1+U+(s-s')\hbar\omega$, where ${\tilde\varepsilon}_1=\varepsilon_1-\lambda_0^2/\hbar\omega$ and
$U$ is the direct interaction term $V_{1111}$ from the two-body Coulomb operator in Eq.\,(\ref{H_S0}). For the parameters 
chosen here we find $U\sim 1.67$ meV. 

One notes that pairs of vibrational components $\{|s_{\nu}\rangle,|s'_{\nu'}\rangle \}$ which differ by the same amount of vibron quanta $\delta$ have equal tunneling 
energies and will therefore contribute simultaneously to the current. However, their Franck-Condon tunneling amplitudes 
are different and decrease if $s,s'$ correspond to excited vibronic states. We also find that the tunneling amplitude of 
the `diagonal' transitions is much larger than the one of the `off-diagonal' transitions (i.e\,for $s\neq s'$) which decreases as 
 $s-s'$ increases.     

\subsection{The turnstile regime}

We denote by $t_p$ the period of the charging/discharging cycles, such that the time needed for each turnstile operation is $2t_p$. 
The value of the loss coefficient $\kappa=0.5\,\mu$eV. The GME was solved numerically on a subspace containing the lowest in energy 20
vibron-dressed states. We have checked that adding more vibronic states will not qualitatively alter the
presented results. Let us mention here that the decreasing value of the FC factors for transitions between highly
excited vibronic states is essential in order to set a reasonably small cutoff $N_0$. In principle one can include more states
in the calculations, but the numerical effort to solve the master equation in the non-markovian regime increases considerably.

\begin{figure}[tbhp!]
\includegraphics[width=0.45\textwidth]{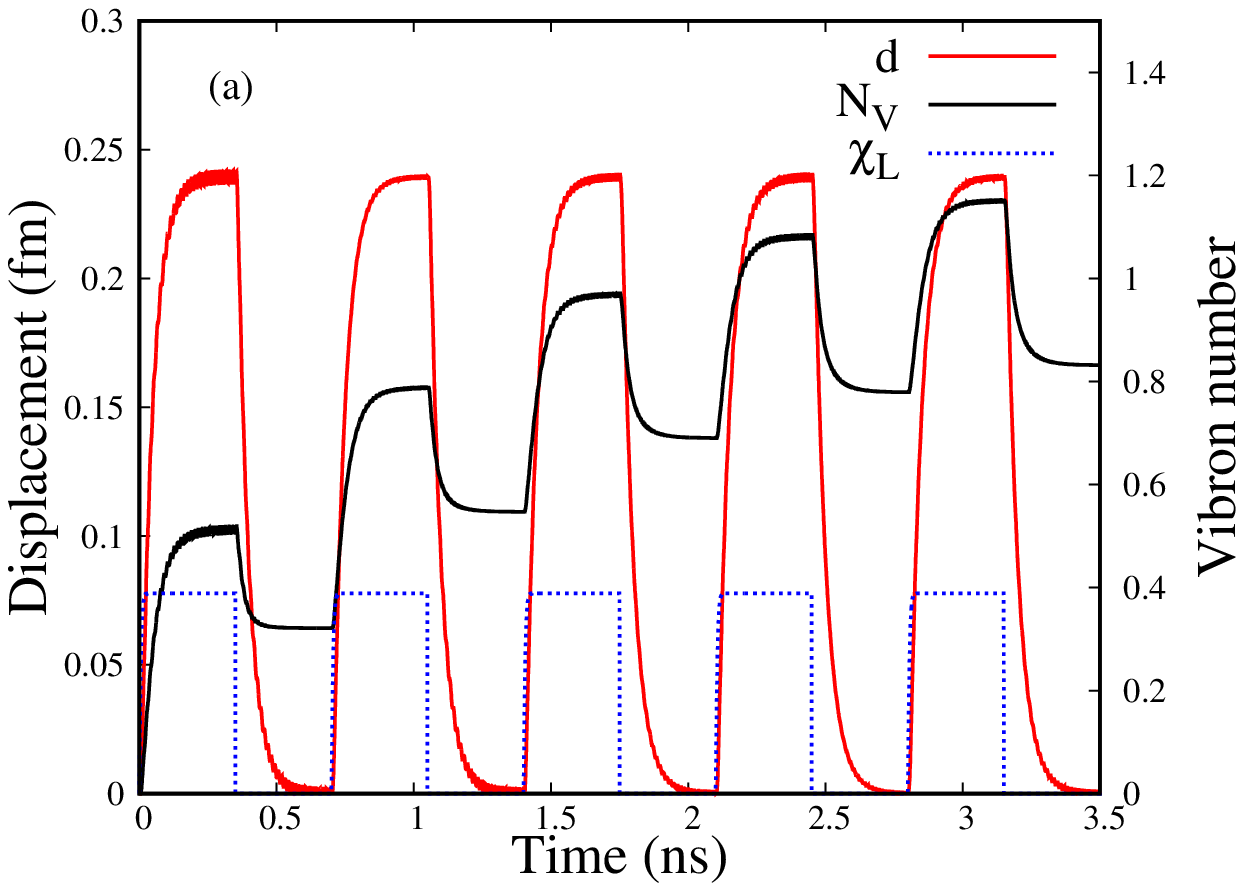}
\includegraphics[width=0.45\textwidth]{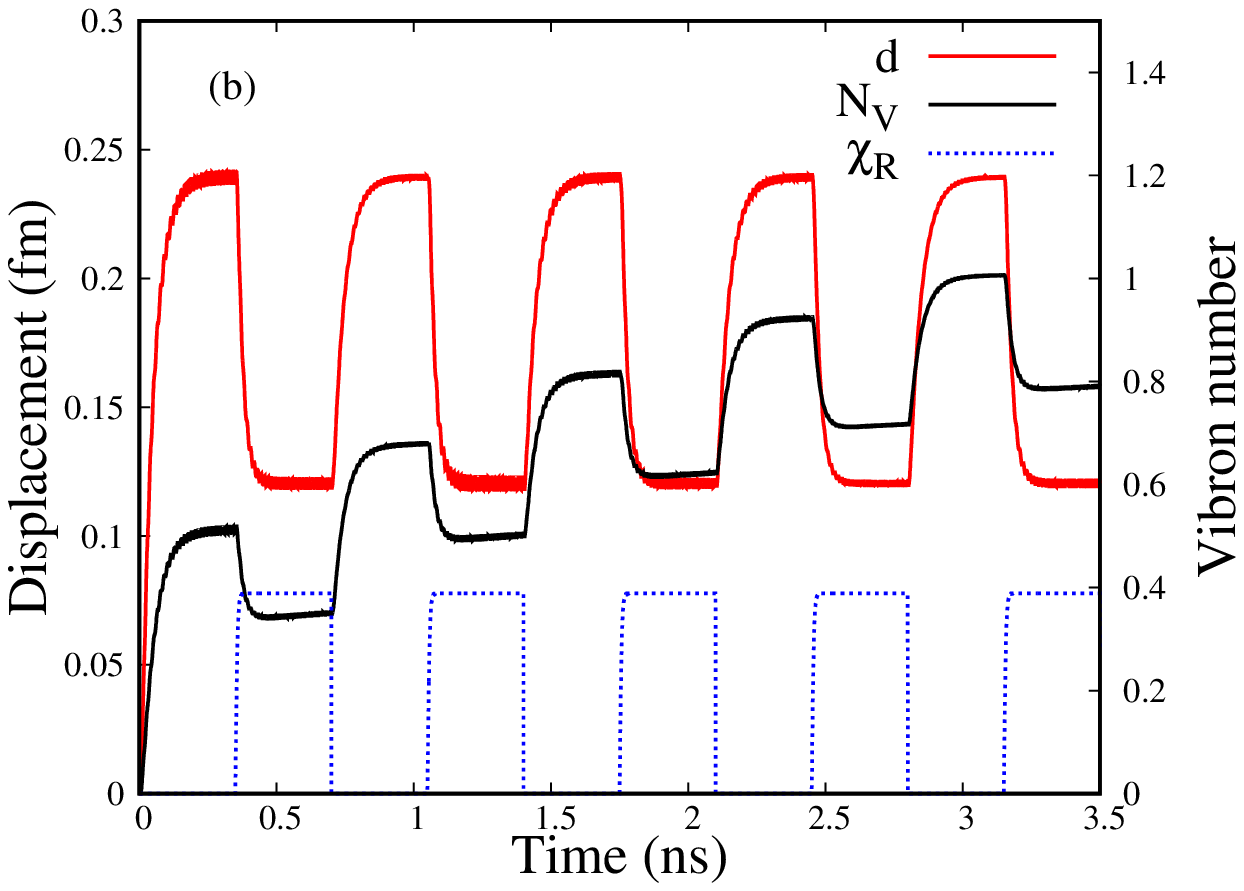}

        \vspace{-1.25cm}

        \includegraphics[width=0.45\textwidth]{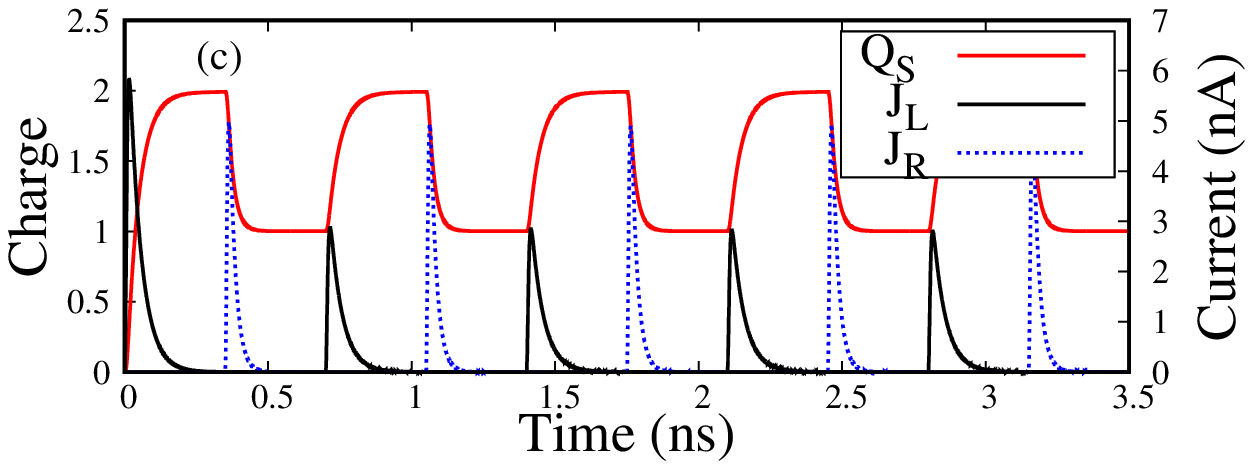}

        \vspace{-1.25cm}

        \caption{The dynamics of the vibron number $N_v$ and the displacement $d$ of the nanoresonator in the two turnstile regimes
which allow the net pumping of $Q$ electrons along each cycle: (a) $Q=2$, $\mu_L=3.5$ meV, $\mu_R=-0.25$ meV and
        (b) $Q=1$, $\mu_L=3.5$ meV, $\mu_R=1.65$ meV. The dotted lines indicate the functions $\chi_{L,R}$ which simulate
        the periodic on and off switching of the two contacts. (c) The charge occupation and the transient currents $J_{L,R}$
        for the $Q=1$ turnstile operation. Other parameters: $t_p=0.35$\,ns .}
\label{fig3-NEMS}
\end{figure}

The periodic switching functions $\chi_{L,R}$ which simulate the turnstile operation are square-shaped and 
oscillate out-of-phase, as shown in Figs.\,\ref{fig3-NEMS}(a) and (b). We assume that the initial state of the hybrid 
system $|\nu=0,N=0\rangle$. The numerical simulations were performed for two turnstile regimes which differ 
by the number of charges $Q$ transferred across the system along each turnstile cycle. In the first regime we set the chemical 
potentials of the leads such that the system is charged with two electrons and then completely depleted, hence $Q=2$. 
For the second regime $\mu_R$ is pushed up to $\mu_R=1.65$\,meV such that the discharging sequence allows only the 
tunneling from the two-particle configuration $|\uparrow\downarrow\rangle$. Then at the end of the turnstile cycle the total 
charge transferred across the system is $Q=1$. The selected values of the chemical potentials for the $Q=1$ and $Q=2$ operations
are also indicated by horizontal solid lines in Fig.\,\ref{fig2-NEMS}. These two regimes should reveal the dependence of the 
electron-vibron coupling on the number of levels contributing to the transport. 

The effects of the turnstile operations $Q=1,2$ on the displacement $d$ and average vibron number $N_v$ are presented in 
Figs.\,\ref{fig3-NEMS} (a) and (b). For the two-particle pumping (see Fig.\,\ref{fig3-NEMS} (a)) the displacement 
of the single-mode nanoresonator roughly mimics the behavior of the potential  $\chi_L$ applied on the left contact. 
More precisely, $d$ increases quickly as the electrons enter the system, saturates once the charge occupation reaches 
the maximum value ${\cal Q}_S=2$ (not shown) and then drops to zero on the discharging half-periods. Note that the oscillations 
of the displacement match the period of the turnstile cycle, $2t_p=0.7$\,ns. It is also clear that the NR bounces 
between a maximum value $d_{{\rm max}}\approx 0.24$\,fm which does not depend on the turnstile cycle and the equilibrium position (i.e.\ $d=0$). 
In particular, we have checked that $d$ and the average charge ${\cal Q}_S$ vanish simultaneously.
This behavior confirms that the electron-vibron coupling is indeed periodically switched on and off along a 
turnstile cycle. 

For $Q=2$ turnstile operation the average vibron number $N_v$ displays a more surprising behavior: (i) It reaches a steady-state 
value $N_v\approx 0.5$ on the first charging sequence but then drops to a lower yet non-vanishing value during the depletion cycle. 
(ii) By repeating the turnstile operation the same pattern is recovered as more vibrons are stored in the NEMS. Eventually, $N_v$ 
reaches a quasi-stationary regime around $t=4.5$\,ns (not shown). We therefore see that along the depletion cycles 
the vibrons are stored in the system in spite of the fact that the electron-vibron coupling is ineffective since ${\cal Q}_S$. 

The single-particle turnstile operation ($Q=1$) leads to a similar behavior of the average vibron number
(see Fig.\,\ref{fig3-NEMS} (b)). However, $N_v$ reaches lower quasistationary values when compared to the
two-particle turnstile operation. A significant  difference is noticed in the displacement oscillations.
At the end of each turnstile cycle the NR does not return to its equilibrium position but settles down to a distance
$d'=d_{{\rm max}}/2$ from its equilibrium position. This happens because in the $Q=1$ turnstile regime the effect of the
electron vibron-coupling is only reduced but not turned off because one electron is always present in the electronic system
and therefore induces a minimal `deflection' of the NR. In this sense, the single-particle turnstile operation can be seen as
a way to dynamically switch between electron-vibron interactions corresponding to a fixed number of particles.
On the other hand, the different response of the NR displacement can be used to `read' the number of charges transferred
across the system along the turnstile cycles, in the presence of vibrons.

In Fig.\,\ref{fig3-NEMS}(c) we plot for completeness the dynamics of the total charge ${\cal Q}_S$ along the single-particle turnstile
operation and the corresponding transient currents $J_{L,R}$. The latter display sharp peaks, their different amplitudes being a
consequence of the different rates at which the system is charged or depleted (note that ${\cal Q}_S$ drops more abruptly on each
discharging half-period).
\begin{figure}[tbhp!]
\vspace{-1.5cm}

\includegraphics[width=0.475\textwidth]{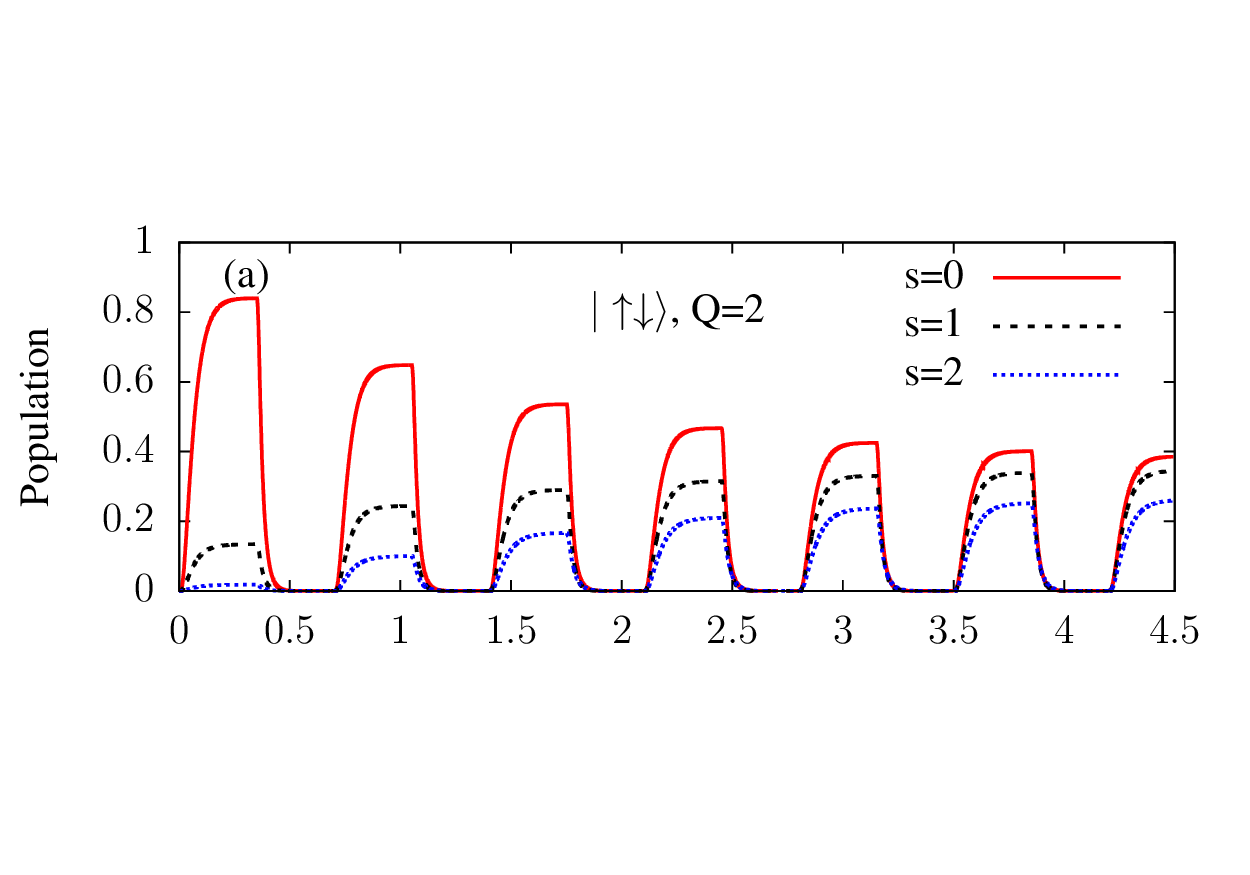}

\vspace{-2.95cm}

\includegraphics[width=0.475\textwidth]{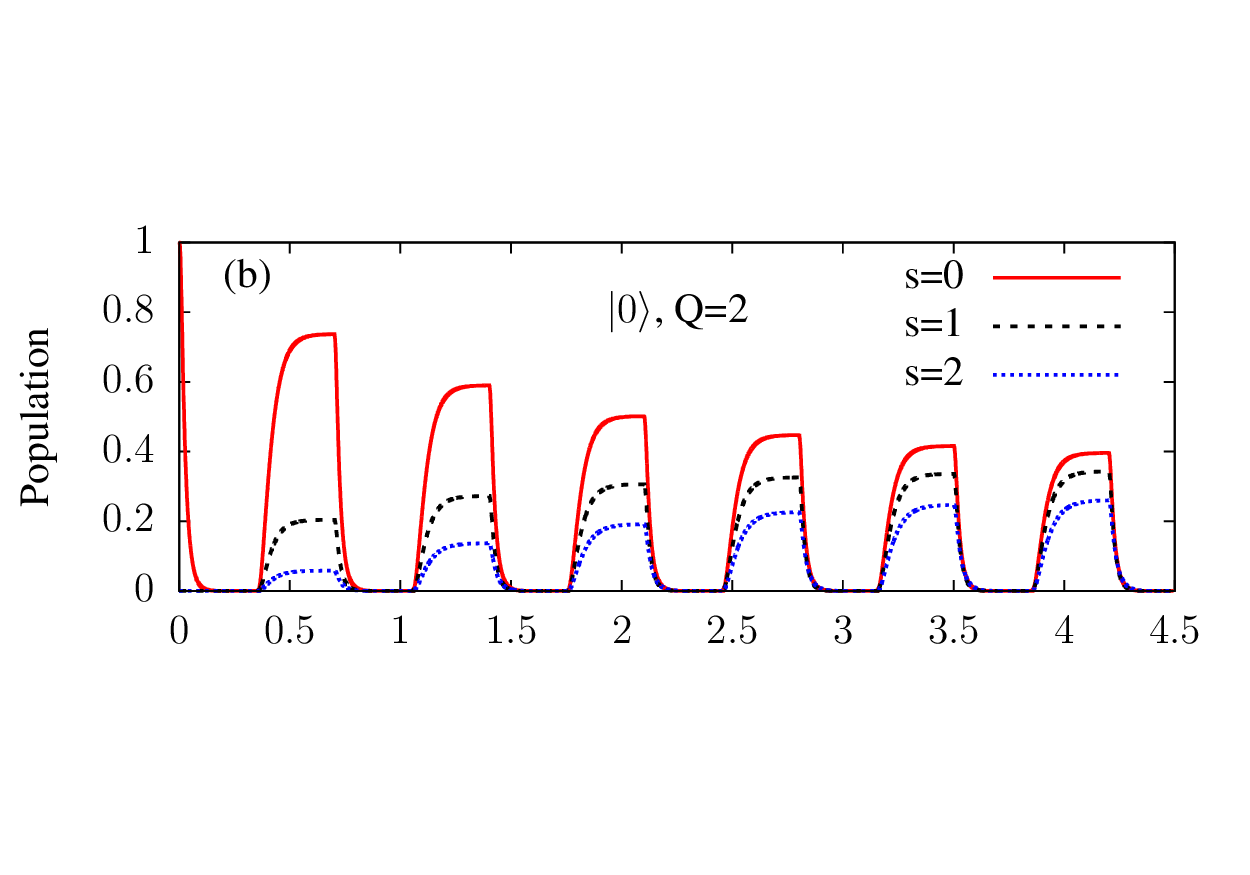}

	\vspace{-2.85cm}

\includegraphics[width=0.475\textwidth]{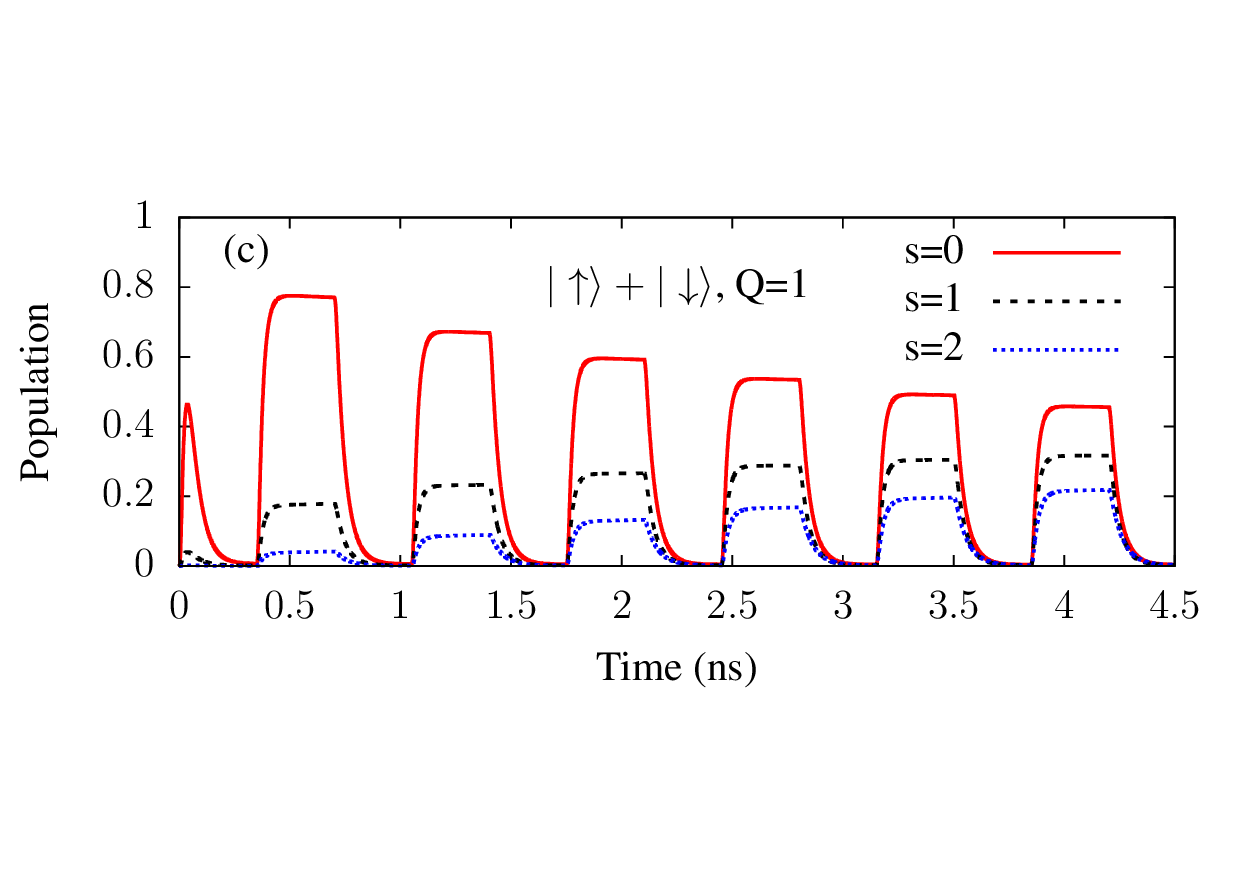}

	\vspace{-1.25cm}

	\caption{The relevant $N$-particle populations $P_{\nu,s}$ of the ground and excited vibronic states: 
	(a)  $P_{\uparrow\downarrow,s}$ for $Q=2$ operation; (b)  $P_{0,s}$ for $Q=2$ operation; 
	(c) The total population of single-particle configurations $P_{1,s}=P_{\uparrow,s}+P_{\downarrow,s}$ for  $Q=1$ operation.}
\label{fig4-NEMS}
\end{figure}
We recall here that an `effective' temperature $T_{{\rm eff}}$ of the hybrid system can be derived from the equilibrium
distribution function $n_B(\omega,T_{{\rm eff}})$ corresponding to the calculated average vibron number $N_v$
(see e.g.\,\cite{Poot}). Using this equivalence we realize that both turnstile
operations induce a sequence of `heating' and {\it partial} `cooling' processes on the NR, as already proved by the vibron dynamics.
To explain this behavior we look more closely at the populations $P_{\nu,s}$ along each turnstile cycle for
$\nu=0,\uparrow,\downarrow,\uparrow\downarrow$. We discuss first the two-electron turnstile operation.
From Fig.\,\ref{fig4-NEMS}(a) we observe that at the end of the charging cycles the hybrid system is completely described
by several two-particle configurations $|\varphi_{\uparrow\downarrow,s}\rangle$ (smaller contributions of $P_{\uparrow\downarrow,s>2}$
were not shown). We also find that the populations $P_{1,s}=\sum_{\sigma}\rho_{\sigma s,\sigma s}$ reach a maximum value shortly after the
coupling of the source lead and then vanish as the two-particle states are filled.

Fig.\,\ref{fig4-NEMS}(b) shows that on the discharging cycles the reduced density matrix of the system 
contains both the `ground' and `excited' purely vibronic states. Moreover, the occupation of the states 
 $|\varphi_{0,s>0}\rangle$ on each depletion half-period increases until a quasistationary regime is reached. 
 This explains why the mean vibron number $N_v$, which collects contributions of the type $w_{\nu,s}P_{\nu,s}$, increases 
 along each turnstile cycle. One can also easily check that the decreasing population of the ground state configuration 
 $|\varphi_{0,s=0}\rangle$ is balanced by the presence of excited vibronic states.  

The accumulation of vibrons in the empty system (i.e., the partial cooling mechanism) can be explained by carefully 
counting the various vibron-assisted tunneling processes connecting pairs of fully interacting states 
$\{\varphi_{\nu,s},\varphi_{\nu',s'}\}$. The chemical potentials of the leads are selected such that all relevant 
tunneling processes (diagonal or off-diagonal) are active, i.e., most of the energies $\Delta_{N,N+1}(s,s')$ are within the bias window 
$(\mu_R,\mu_L),$ for $N=0,1$ (see the chemical potentials for the $Q=2$ setting in Fig.\,\ref{fig2-NEMS}). 
For the $Q=1$ operation we have instead $\Delta_{0,1}(s-s')<\mu_R<\Delta_{1,2}(s-s')<\mu_L$ for the most important tunneling processes.  
When looking at Fig.\,\ref{fig2-NEMS} we notice that some transitions are left outside the bias window, e.g.\,the ones corresponding to
$\Delta_{0,1}(\delta=3,4)$. However, these transitions have a small tunneling amplitude and they will not significantly contribute to 
the transport. It is easy to see that for the first charging cycle of the $Q=2$ operation the sequence of `diagonal' transitions 
e.g\,$|\varphi_{0,0}\rangle\to|\varphi_{\sigma=\uparrow,0}\rangle\to|\varphi_{\uparrow\downarrow,0}\rangle$ involves 
only the lowest vibronic components ($s=s'=0$) with small vibron numbers $w_{\nu,0}$ (see Eq.\,(\ref{vib-num})). These transitions are also 
the strongest, as the corresponding FC factors are the largest ones. Note also that on the first charging cycle the vibron absorption 
is not possible as the initial state is $|\varphi_{0,0}\rangle$, such that the `excited' states $|\varphi_{\sigma,s>0}\rangle$ can only 
be populated through  `off-diagonal' weaker transitions, for example $|\varphi_{0,0}\rangle\to|\varphi_{\sigma,1}\rangle$.
Finally, the charging cycle brings the second electron to the system and opens more tunneling paths involving both diagonal and 
off-diagonal processes, e.g., $|\varphi_{0,0}\rangle\to|\varphi_{\sigma,1}\rangle\to|\varphi_{\uparrow\downarrow,1}\rangle$ or 
$|\varphi_{0,0}\rangle\to|\varphi_{\sigma,1}\rangle\to|\varphi_{\uparrow\downarrow,2}\rangle$. 
It is therefore clear that at the end of the first charging cycle the system is described by the two-particle electronic 
configuration $|\uparrow\downarrow\rangle$ and several vibronic components $|s_{\uparrow\downarrow}\rangle$ with the 
associated vibron numbers $w_{\uparrow\downarrow,s}$. 

Now, during the first depletion cycle this mixed structure of the reduced density matrix allows the activation of multiple 
`diagonal' and `off-diagonal' tunneling out processes between $(N+1)-$particle and $N-$particle configurations. For example the 
`diagonal' backward sequence $|\varphi_{\uparrow\downarrow,1}\rangle\to|\varphi_{\sigma,1}\rangle\to|\varphi_{0,1}\rangle$ 
leaves the hybrid system in the first vibronic excited state whose population $P_{0,1}\approx 0.2$ in Fig.\,\ref{fig4-NEMS}(b).   
The small population $P_{0,2}$ is due to the similar `off-diagonal' transition from 
$|\varphi_{\sigma,1}\rangle\to|\varphi_{0,2}\rangle$. 
Other transitions leading to vibrational `cooling' can be also identified. As a result the mean vibron
number drops over the depletion cycle, but does not vanish due to the `diagonal' tunneling events. 

At the next charging cycle the excited single-particle states $|\varphi_{\sigma,1}\rangle$ will be fed by both diagonal 
and off-diagonal transitions, because when switching on the coupling to the left lead the reduced density matrix of the 
system reads $\rho(2t_p)=\sum_s|\varphi_{0,s}\rangle\langle\varphi_{0,s}|$. As a consequence, the population $P_{\uparrow\downarrow,1}$ 
almost doubles with respect to the first charging cycle, whereas $P_{\uparrow\downarrow,2}$ brings a small contribution as well. 
The vanishing of the displacement $d$ on each discharging sequence 
is mandatory, as the system is completely described by purely vibronic states and therefore 
$\langle\varphi_{0,s}|a^{\dagger}|\varphi_{0,s}\rangle=0$.

In Fig.\,\ref{fig4-NEMS}(c) we present for completeness the populations of one-particle configurations which 
describe the hybrid system for the $Q=1$ turnstile operation. Clearly, the discharging cycles are now described by 
single-particle states $|\varphi_{\sigma=\uparrow,\downarrow,s}\rangle$. The empty states $|\varphi_{0,s}\rangle$ 
are no longer accessible in this case so they were not shown. By comparing Figs.\,\ref{fig4-NEMS} (b) and (c) one notices  
a lower occupation of the excited states $|\varphi_{\sigma=\uparrow,\downarrow,s>0}\rangle$ which explains why the `jumps' 
and drops of the mean vibron number are less pronounced. This could be expected because the electron-vibron coupling 
is now enhanced/reduced only due to a single electron which is added/removed from the system.       

\begin{figure}[tbhp!]
\includegraphics[width=0.45\textwidth]{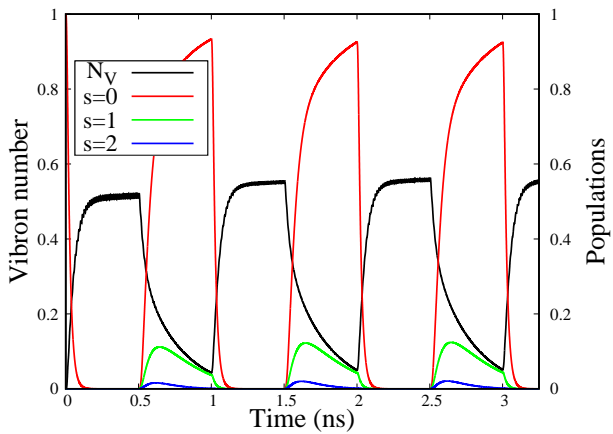}
        \caption{The dynamics of vibron number $N_v$ and of the populations $P_{0,s}$ for the
        the $Q=2$ turnstile protocol. In contrast to Fig.\,\ref{fig3-NEMS}(a) the complete `cooling' of the nanoresonator
is insured by suppressing the tunneling-out tunneling processes. Other parameters: $\mu_L=3.5$\, meV, $\mu_R=0.68$\, meV, $t_p=0.5$\, ns.}
\label{fig5-NEMS}
\end{figure}

In the following we investigate in more detail the role of the bias window on the partial cooling processes in the $Q=2$ turnstile operation.
To this end the chemical potential of the drain reservoir is pushed up to $\mu_R=0.68$\,meV such that the main `heating' processes 
associated to the depletion cycles are forbidden, that is $\Delta_{0,1}(s,s')<\mu_R$ for some $s<s'$ (see the lowest dotted horizontal 
line in Fig.\,\ref{fig2-NEMS}). Figure\,\ref{fig5-NEMS} shows that in this case the mean vibron number does not display steps on the discharging 
cycles (as in Fig.\,\ref{fig3-NEMS}(b)) but rather vanishes - in other words, the hybrid system eventually cools down to the temperature 
of the thermal bath $T$. In order to capture the slow evolution of $N_v$ we increased the turnstile period to $t_p=1$\,ns. Further insight into the 
vibron dynamics is given by the populations  $P_{0,s}$ of the purely vibronic states which are also presented in 
Fig.\,\ref{fig5-NEMS}. 
After an initial increase, the excited states $|\varphi_{0,1}\rangle$ and $|\varphi_{0,2}\rangle$ are slowly depleted in 
favor of the ground state $|\varphi_{0,0}\rangle$ whose population increases uniformly on each discharging sequence. 
This behavior differs from the one shown in Fig.\,\ref{fig4-NEMS}(b) and suggests a `redistribution' of probability between various 
purely vibronic states. In the following we explain this 
effect through the interplay of tunneling-out and -in processes which involve the drain lead. 

The sudden drop of $N_v$ right after opening the contact to the right reservoir is due to the `cooling' transitions 
$|\varphi_{\sigma,s}\rangle\to |\varphi_{0,s'}\rangle$ for $\delta=s-s'>0$, whose energies are still 
above $\mu_R$ (see Fig.\,\ref{fig2-NEMS}). On the other hand, the excited vibronic states $|\varphi_{\sigma,s>0}\rangle$ 
are still being populated via vibron-conserving transitions $|\varphi_{\sigma,s}\rangle\to |\varphi_{0,s}\rangle$ and to a 
lesser extent by the partial `cooling' transition $|\varphi_{\sigma,2}\rangle\to |\varphi_{0,1}\rangle$. This scenario is 
confirmed by the initial increase of the populations $P_{0,1}$ and $P_{0,2}$. We find instead that the much slower vibronic 
relaxation involves two more sequential tunnelings, one from the reservoir to the central system and another one back to it. 
Indeed, given the fact that $\Delta_{0,1}(\delta<0)$ are below $\mu_R$, electrons can tunnel {\it back} from the contact via 
`cooling' transitions $|\varphi_{0,s}\rangle\to |\varphi_{\sigma,s'}\rangle$ (for $s'<s$). Finally, the lower-temperature 
single-particle states are depleted through diagonal transitions $|\varphi_{\sigma,s'}\rangle\to|\varphi_{0,s'}\rangle$. 

Turning back to the symmetric bias setting (see Fig.\,\ref{fig3-NEMS} (a)), it is readily seen that the tunneling-mediated 
cooling mechanism presented above cannot be active. In this case, electrons are not allowed to tunnel back to the central 
system because $\mu_R$ lies below all transition energies. Moreover, the cooling processes 
$|\varphi_{\sigma,s'}\rangle\to |\varphi_{0,s}\rangle$ with $s<s'$ are overcome by the heating processes such that 
$N_v$ settles down to a non-vanishing value after the onset of the discharging sequence.

\begin{figure}[tbhp!]
\includegraphics[width=0.45\textwidth]{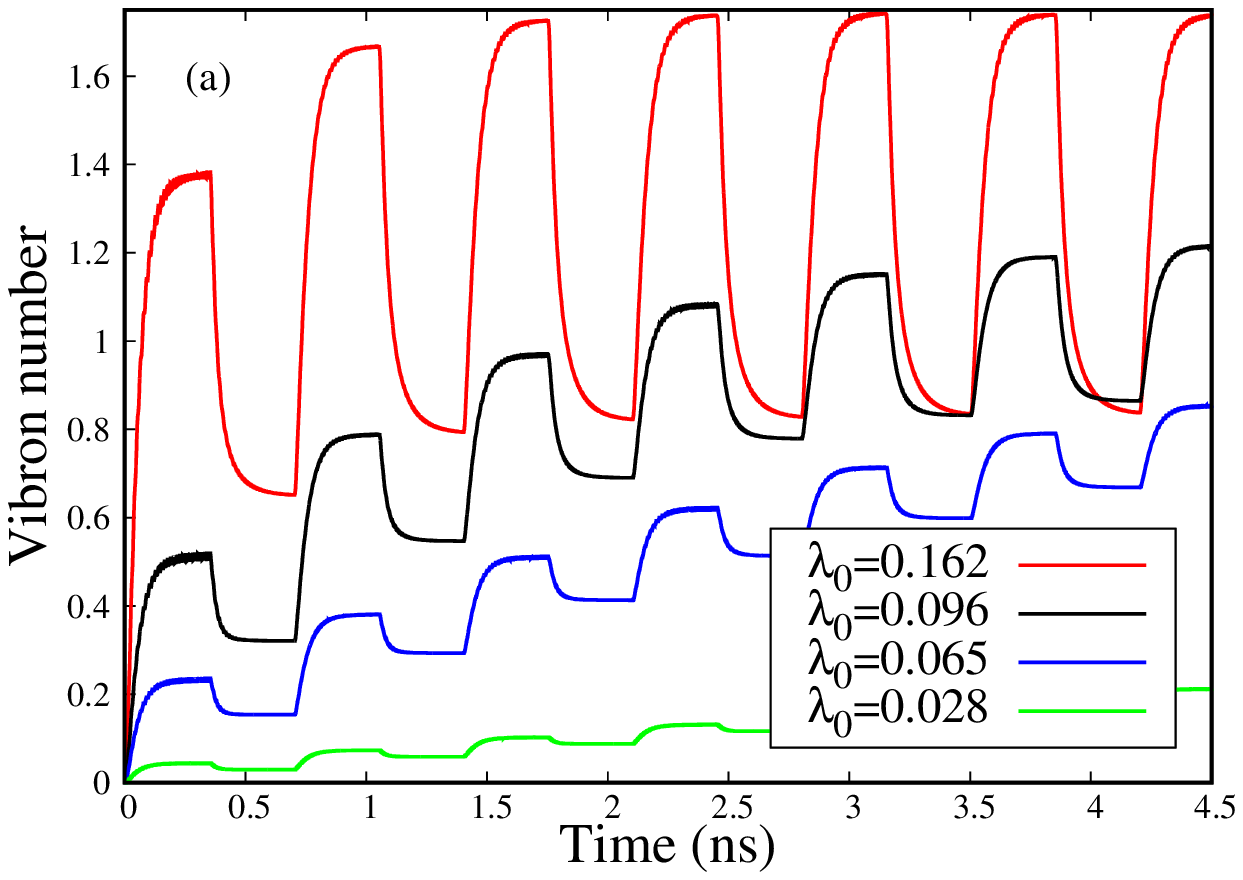}
\includegraphics[width=0.45\textwidth]{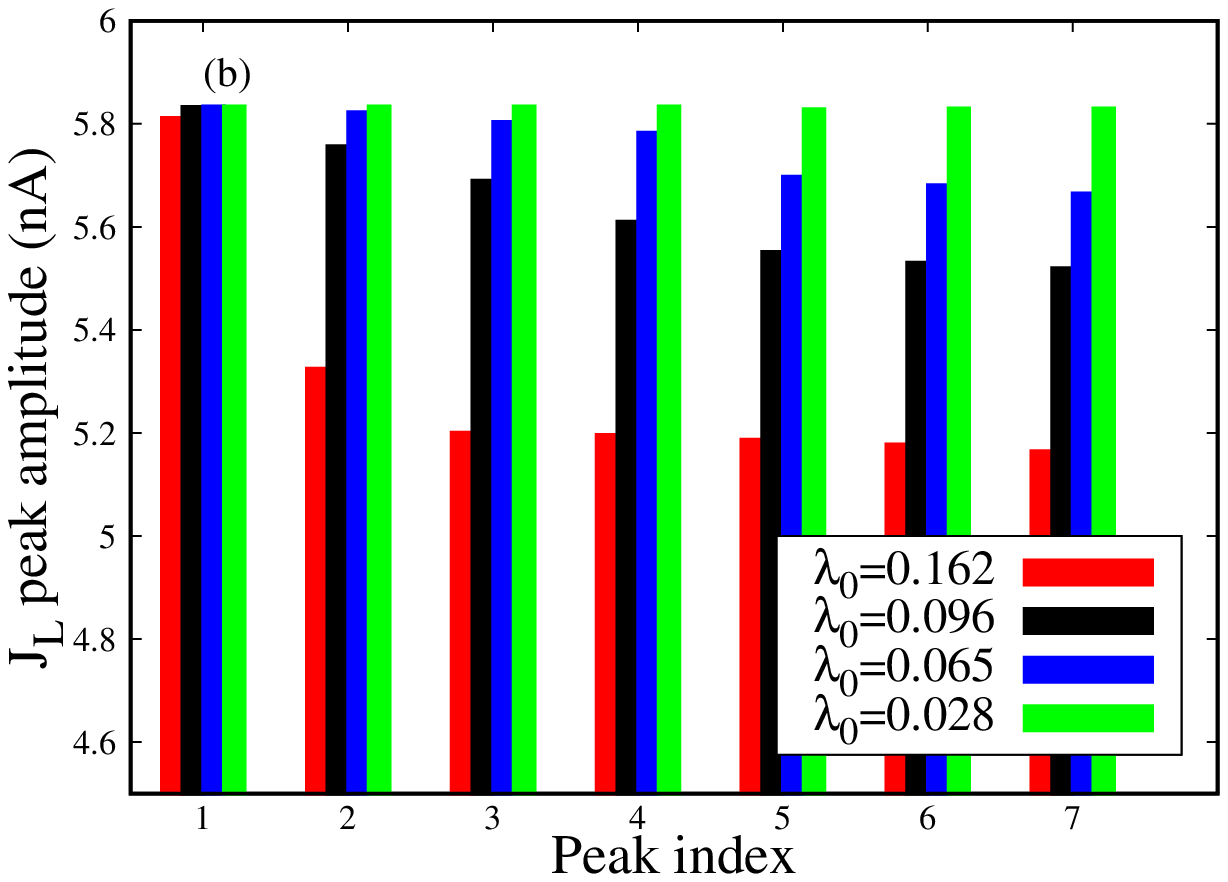}
	\caption{The effect of the electromechanical coupling strength $\lambda_0$ (given in meV units) on (a) the vibron number $N_v$
and (b) on the peak amplitude of the transient current $J_L$. The parameters correspond to the 
$Q=2$ turnstile protocol: $\mu_L=3.5$\, meV, $\mu_R=-0.25$\, meV, $t_p=0.35$\, ns.}
\label{fig6-NEMS}
\end{figure}

In order to check whether the electron-vibron coupling affects not only the dynamics of the NR but 
also the transport properties of the electronic subsystem, we present in Fig.\,\ref{fig6-NEMS}(a) the 
vibron dynamics for several values of the electron-vibron coupling strength $\lambda_0$. This parameter can 
be tuned by changing either the equilibrium distance between the electronic system and the
nanoresonator (as shown in previous work \cite{NEMS-1}) or the NR mass $M$. The amplitude of the
heating and cooling cycles decreases with $\lambda_0$ and the hybrid system approaches the quasistationary
regime much faster at larger values of $\lambda_0$. For example, a considerable difference is noticed between
the first two cycles at $\lambda_0=0.162$ meV, the next cycles being rather similar.

In Fig.\,\ref{fig6-NEMS}(b) we collect the amplitudes associated to the first seven peaks of the current $J_L$ 
and to the different electron-vibron couplings considered in Fig.\,\ref{fig6-NEMS}(a). The peak evolution over few 
turnstile cycles can also be extracted from transport measurements and provides indirect insight on the vibron 
dynamics. In the weakly interacting case ($\lambda_0=0.028$ meV) the amplitudes of the peaks are nearly 
equal and one cannot discern the negligible effect of the electron-vibron coupling on the transport properties. 
In contrast, as $\lambda_0$ increases, the peaks display noticeable differences. More precisely, their amplitude 
gradually decreases from one cycle to another until it reaches a quasistationary value (for $\lambda_0=0.096$ meV this
value is roughly 5.5 nA). Note that the first peak of the charging current $J_L$ is less sensitive w.r.t. changes 
of $\lambda_0$ because at such short times the vibrons are not yet activated. For the larger value $\lambda_0=0.162$ 
meV a steep reduction of the peak is noticed after two charging half-periods. A similar behavior is recovered for 
the output current $J_R$ (not shown). By comparing Figs.\,\ref{fig6-NEMS}(a) and (b) one infers that the attenuation 
of the peak amplitude is correlated to the emergence of the quasistationary regime for the heating/cooling sequences.

\begin{figure}[t]
\includegraphics[width=0.45\textwidth]{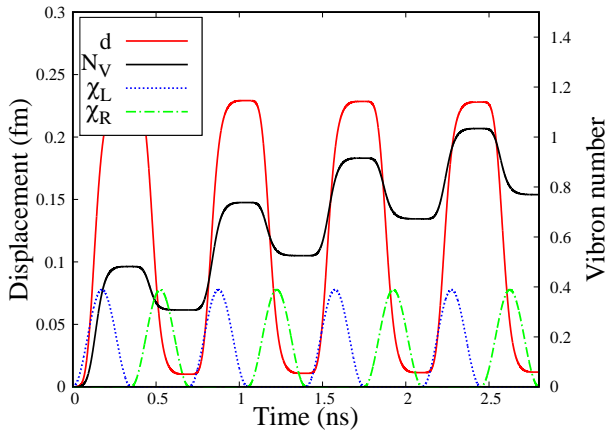}
        \caption{The dynamics of the vibron number $N_v$ and displacement $d$ of the nanoresonator for smoother switching
	functions $\chi_{L,R}$. Other parameters: $\lambda_0=0.096$\,meV, $\mu_L=3.5$\, meV, $\mu_R=-0.25$\, meV, $t_p=0.35$\, ns.}
\label{fig7-NEMS}
\end{figure}

We also considered other shapes for the switching functions $\chi_{L,R}$ and we recovered similar effects of 
the turnstile regime on the nanoresonator, i.e. heating/cooling on the charging/discharging half-periods. 
 Figure\,\ref{fig7-NEMS} shows the vibron dynamics $N_v$ and the displacement $d$ for smoother switching functions. 
 When compared to the results presented in Fig.\,3(a) we noticed minor changes in the local maximum and minimum values 
of the average vibron number. However, the most important effect is a delay of nanoresonator's response to 
the switching functions, i.e. $N_v$ and $d$ do not increase/decrease immediately after charging/discharging.
If one is interested in implementing faster heating and cooling processes separated by longer `isotherms'
(i.e\, time intervals with constant vibron number $N_v$) the square-wave driving is the most effective.

Finally, we stress that  the oscillations of the displacement record the charge variations along the
turnstile operations but do not discern between the vibron dynamics. In order to understand why this happens let us observe first 
that if the coherences $\langle\varphi_{\nu,s}|\rho(t)|\varphi_{\nu,s'}\rangle$ 
are negligible then from Eq.\,(\ref{displ-diag}) one gets a simpler formula for the displacement:
\begin{equation}\label{displ1}
        d\approx \frac{2\lambda_0 l_0}{\hbar\omega}\sum_{\nu,s}n_{\nu}P_{\nu,s}.
\end{equation}
Secondly, since on the charging sequences the system settles down to the two-electron configuration (i.e., $n_{\nu}=2$) and 
$\sum_sP_{\uparrow\downarrow,s}=1$ for all chemical potentials $\mu_R<\Delta_{0,1}(s-s'=0)$ it follows that $d$ 
cannot depend on $\mu_R$, even if each occupation $P_{\nu,s}$ does. Eq.\,(\ref{displ1}) also confirms the doubling of 
the quasistationary displacement $d_{{\rm max}}$ on the charging cycles with respect to the value attained along the 
depletion cycles of the $Q=1$ turnstile operation, as shown in Figs.\,\ref{fig3-NEMS}(a) and (b).

For the parameters selected here the coherences corresponding to states with the same electronic configurations
but different vibron numbers (i.e., $\langle\varphi_{\nu,s}|\rho(t)|\varphi_{\nu,s'}\rangle$) do exist but they are indeed too small
to induce a noticeable change of the various observables (not shown). In fact, we record some fast oscillations of the displacement
on the `steps' of each turnstile cycle; the period of these oscillations coincides with those of the coherences mentioned
above but one can see from Fig.\,\ref{fig3-NEMS} that their amplitude is hardly noticeable.

Based on these results we state that the quantum turnstile regime provides a dynamical switching of the electron-vibron coupling
effects on the hybrid system. Once the depletion process is complete the electron-vibron coupling is ineffective. However, the 
effect of the latter is imprinted in the non-vanishing populations of the excited vibrational states $\varphi_{0,s>0}$. 
Alternatively, by pumping one electron per turnstile cycle while keeping the lowest level occupied one initializes a configuration 
made by single-particle states 'dressed' by vibrons.

On the other hand, the charging cycles activate the electrostatic coupling and the vibron number increases. It also turns out that
both turnstile operations induce a heating of the nanoresonator when the electron-vibron
coupling is turned on and at least a {\it partial} cooling when it is turned off. 

\section{Conclusions}

We proposed and studied theoretically a quantum turnstile protocol for switching on and off the effect of electron-vibron coupling 
between a biased mesoscopic system and a vibrational mode. A detailed analysis of the vibron-assisted tunneling processes is provided 
by the populations of the vibron-dressed states which are calculated within the generalized master equation method. We identify the 
role of various tunneling processes in the vibron emission (heating) and absorption (cooling) processes. 
The turnstile charging and discharging cycles impose periodic variations of the nanoresonator's displacement with respect to its 
equilibrium value. As the electronic system empties the displacement vanishes. Instead, a turnstile operation which allows only a 
partial depletion sets a lower bound of the displacement due to an extra electron residing in the system. 

The values of the displacement 
obtained in our model are probably too small to be detected. However, $d$ increases as 
more electrons tunnel across the system during a turnstile cycle. This could be achieved by increasing the bias window 
such that more electronic configurations participate in transport. Alternatively, one can consider lighter nanoresonators
and therefore larger values of the oscillator length $l_0$.

We find that in general the average number of vibrons does not vanish along the discharging cycles when the electron-vibron coupling 
is ineffective. In the quasistationary regime the same amount of vibrons is emitted and absorbed along a turnstile cycle. 
Otherwise stated, the system undergoes periodic heating and cooling processes. A complete cooling to the equilibrium temperature of the 
leads or of a thermal bath can be achieved by a suitable choice of the chemical potential of the drain reservoir. 
We also show that the peak amplitude of the transient currents decreases as the strength of the electron-vibron coupling 
increases. Moreover, it turns out that as the heating/cooling cycles attain the quasistationary regime the peak amplitude 
gradually reduces to a value which does not depend on the charging/discharging half-period.

Let us emphasize that the quantum turnstile dynamics differs considerably from the normal transport regime when both leads 
are simultaneously coupled to the system and for which one can only notice a heating process, as the average vibron number
uniformly increases before reaching its stationary value. Also, in the present setting the actuation of the nanoresonator 
is only due to the electronic current as there is no additional driving signal. In other words, we consider that before 
the electronic subsystem is coupled to the leads the nanoresonator is in the static deflection mode.

\begin{acknowledgments}
R.D., V.M. and S.S. acknowledge financial support from CNCS - UEFISCDI grant PN-III-P4-ID-PCE-2016-0221
and from the Romanian Core Program PN19-03 (contract No. 21 N/08.02.2019). B.T. and V.M. were also 
supported by TUBITAK Grant No. 117F125. B.T. further acknowledges the support from TUBA.
\end{acknowledgments}

\end{document}